\documentclass[a4paper,fleqn]{cas-sc}
\usepackage{soul}
\usepackage[sort&compress,numbers]{natbib}
\usepackage{lineno}
\usepackage{setspace}
\usepackage{verbatim}
\usepackage{graphics}
\usepackage{graphicx}
\usepackage{cuted}
\usepackage{lipsum}
\usepackage{multirow}
\usepackage[version=4]{mhchem}
\usepackage[utf8]{inputenc}
\usepackage{siunitx}
\usepackage{gensymb}
\usepackage{hyperref}
\usepackage{blindtext}
\usepackage{comment}
\usepackage{amsmath}
\usepackage{siunitx}
\DeclareSIUnit\angstrom{\text{Å}}
\newcommand{\angstrom}{\mbox{\normalfont\AA}}
\usepackage{url} 
\usepackage{hyperref}
\usepackage{xurl} 
\DeclareUnicodeCharacter{03B2}{\(\beta\)}

\def\tsc#1{\csdef{#1}{\textsc{\lowercase{#1}}\xspace}}
\tsc{WGM}
\tsc{QE}
\tsc{EP}
\tsc{PMS}
\tsc{BEC}
\tsc{DE}

\shorttitle{Sodium-Decorated Ennea-Graphene}
\shortauthors{Bill D. Aparicio-Huacarpuma \textit{et~al}.}

\begin{document}

\title [mode = title]{Sodium-Decorated Ennea-Graphene: A Novel 2D Carbon Allotrope for High-Capacity Hydrogen Storage}

\author[1,2]{Bill D. Aparicio Huacarpuma}
\cormark[1-2]
\ead{bdaparicioh@gmail.com}
\author[3]{José A. S. Laranjeira}
\author[3]{Nicolas F. Martins}
\author[3]{Julio R. Sambrano}
\author[4]{F\'{a}bio L. Lopes de Mendonça}
\author[6]{Alexandre C. Dias}
\author[1,2]{Luiz A. Ribeiro Junior}

\address[1]{\small Institute of Physics, University of Bras{\'{i}}lia, 70919-970, Bras{\'{i}}lia, DF, Brazil.}
\address[2]{\small Computational Materials Laboratory, LCCMat, Institute of Physics, University of Bras{\'{i}}lia, 70919-970, Bras{\'{i}}lia, DF, Brazil.}
\address[3]{\small Modeling and Molecular Simulation Group, São Paulo State University (UNESP), School of Sciences, Bauru, 17033-360, SP, Brazil.}
\address[4]{\small College of Technology, Department of Electrical Engineering, University of Bras\'{i}lia, 70910-900, Bras\'{i}lia, Federal District, Brazil.}
\address[5]{\small University of Bras\'{i}lia, College of Technology, Department of Mechanical Engineering, 70910-900, Bras\'{i}lia, Federal District, Brazil.}
\address[6]{\small Institute of Physics and International Center of Physics, University of Bras{\'{i}}lia, 70919-970, Bras{\'{i}}lia, DF, Brazil.}

\begin{abstract}
The development of safe, efficient, and reversible hydrogen storage materials is critical for advancing hydrogen-based energy technologies and achieving carbon-neutral goals. Ennea-Graphene, a new 2D carbon allotrope made of 4-, 5-, 6-, and mainly 9-membered carbon rings (nonagons), is introduced via Density Functional Theory calculations. Phonon dispersion and \textit{ab initio} molecular dynamics demonstrate that the monolayer is mechanically and dynamically stable at \SI{300}{\kelvin}, as no imaginary modes are detected. The pristine system further exhibits metallic-like electronic behavior. The material exhibits high in-plane stiffness (Young’s modulus $\approx$ \SI{255}{N/m}). Sodium adsorption at the centers of the nonagonal rings is energetically favorable, with a binding energy of approximately -1.56~\si {\electronvolt}, leading to the formation of the Na@Ennea-graphene complex. The calculated \ce{H2} adsorption energies range from \SIrange{-0.15}{-0.18}{\electronvolt}. The \ce{Na}-decorated structure demonstrates excellent hydrogen storage performance, reversibly adsorbing up to four \ce{H2} molecules per Na atom ($\sim$8.8~wt\si{\percent} \ce{H2}). This capacity surpasses the U.S. Department of Energy’s 2025 target for onboard hydrogen storage materials. The adsorbed \ce{H2} remains molecular (\ce{H}--\ce{H} bond $\sim$\SI{0.76}{\angstrom}) and can be released under near-ambient conditions, as verified by \SI{300}{\kelvin} \textit{ab initio} molecular dynamics simulations. These findings position sodium-decorated Ennea-Graphene as a promising nanomaterial for next-generation hydrogen storage technologies.
\end{abstract}




\maketitle

\doublespacing

\section{Introduction}

The transition toward a carbon-neutral energy economy requires the development of safe, efficient, and reversible hydrogen storage systems.\cite{guilbert2021hydrogen, cheekatamarla2024hydrogen} Hydrogen is especially attractive because of its high gravimetric energy density, making it a strong candidate for fuel cell vehicles and stationary storage.\cite{sadeq2024hydrogen, tashie2021hydrogen} Despite these advantages, the lack of materials that combine high capacity, stability, and reversibility under near-ambient conditions remains a central obstacle to large-scale implementation.\cite{sun2023brief, ma2024large, heinemann2021enabling}  

To guide research in this area, the U.S. Department of Energy (DOE) has established system-level targets for onboard hydrogen storage, including a 2025 benchmark of 6.5~wt\si{\percent}.\cite{DOE2025HydrogenStorageTargets} These criteria have intensified efforts to design nanostructured systems that can meet capacity requirements while remaining stable and reversible in realistic operating windows. Computational methods, particularly density functional theory (DFT), have played a central role in accelerating this search, enabling reliable predictions of stability and performance prior to experimental realization.\cite{CHEN2024510, LIU2025105802, djebablia2024metal,ABDULLAHI2025116631}  

Within this context, two-dimensional (2D) carbon allotropes have emerged as highly promising candidates.\cite{abifarin20242d, ghotia2025review} Their low atomic mass, large surface areas, and versatile bonding motifs provide a tunable platform for hydrogen adsorption. Considerable attention has been directed to graphene,\cite{kag2021strain, morse2021hydrogenated} biphenylene networks,\cite{ma2024li, chotsawat2024first} graphyne,\cite{liu2014hydrogen, guo2013comparative, zhang2025cli3} graphdiyne,\cite{jiang2023density, bajgirani2024boosting} and other exotic lattices.\cite{laranjeira2025tphe, laranjeira2025potassium, laranjeira2025oli3, vaidyanathan2025strain, wei2025rc14} A widely used strategy to enhance hydrogen uptake is the decoration of these lattices with light metals, which strengthen the \ce{H2} binding through polarization and spillover effects.\cite{mohajeri2018light, rowsell2005strategies}  

Building upon these advances, we report a first-principles investigation of Ennea-Graphene, a novel 2D carbon allotrope characterized by a periodic arrangement of 4-, 5-, 6-, and 9-membered rings. This structure was generated computationally using Crystal-RG2, a stochastic design code combining group and graph theory.\cite{shi2021high,gong2020theoretical,yin2019stone,he2025isolated} Structural, mechanical, electronic, and thermal stability were examined, along with hydrogen storage properties under sodium decoration. Our results reveal that Ennea-Graphene not only exhibits excellent stability but also achieves a reversible hydrogen storage capacity of 8.8~wt\si{\percent}, surpassing the DOE 2025 target. This work introduces an unexplored carbon architecture and provides insights that may inspire both theoretical studies and experimental synthesis of 2D carbons with rare ring topologies.

\section{Computational Methods}

The calculations were performed within the DFT framework using the Vienna \textit{ab initio} simulation package (VASP) \cite{Kresse_13115_1993,Kresse_11169_1996}. The generalized gradient approximation (GGA) \cite{Perdew1996} with the Perdew--Burke--Ernzerhof (PBE) functional \cite{perdew_1_1991} was employed to describe exchange-correlation effects, combined with the projector-augmented wave (PAW) \cite{Blchl1994} method. A plane-wave cutoff energy of \SI{550}{\electronvolt} was used for all calculations to ensure convergence of the total energy. The Brillouin zone was sampled using $\Gamma$-centered \textbf{k}-point grids with a $5\times5\times1$ mesh for structural optimization, and a denser $10\times10\times1$ mesh for electronic structure analysis. A vacuum layer of \SI{15}{\angstrom} was added perpendicular to the 2D sheet to prevent spurious interactions between its periodic images.

Structural relaxation was performed until the total energy converged to within \SI{E-6}{\electronvolt} and the residual forces on each atom were smaller than \SI{0.01}{\electronvolt/\angstrom}. To address the well-known underestimation of band gaps by the PBE functional, the electronic band structure calculations were also performed using the range-separated hybrid exchange-correlation functional proposed by Heyd--Scuseria--Ernzerhof (HSE06).\cite{Heyd2006} Phonon dispersion relations were obtained to assess dynamical stability using the PHONOPY package,\cite{Togo2023}, and elastic constants were computed to evaluate mechanical robustness. Dispersion interactions were included using the DFT-D2 method of Grimme.\cite{Grimme2006}

AIMD simulations were carried out in the canonical (NVT) ensemble using a Nosé-Hoover thermostat \cite{Hoover1985} to maintain the system at \SI{300}{\kelvin}. A time step of \SI{0.5}{\fs} was used, and simulations were run for a total duration of \SI{5}{\ps} to monitor the thermal stability of both pristine and sodium-decorated structures. Adsorption energies of sodium and hydrogen molecules were computed by comparing the total energy of the adsorbed systems with those of isolated components. 

\section{Results and Discussion}

\subsection{Structural and Dynamical Stability of Ennea-Graphene}

\begin{figure}[ht]
\centering
\includegraphics[width=\textwidth]{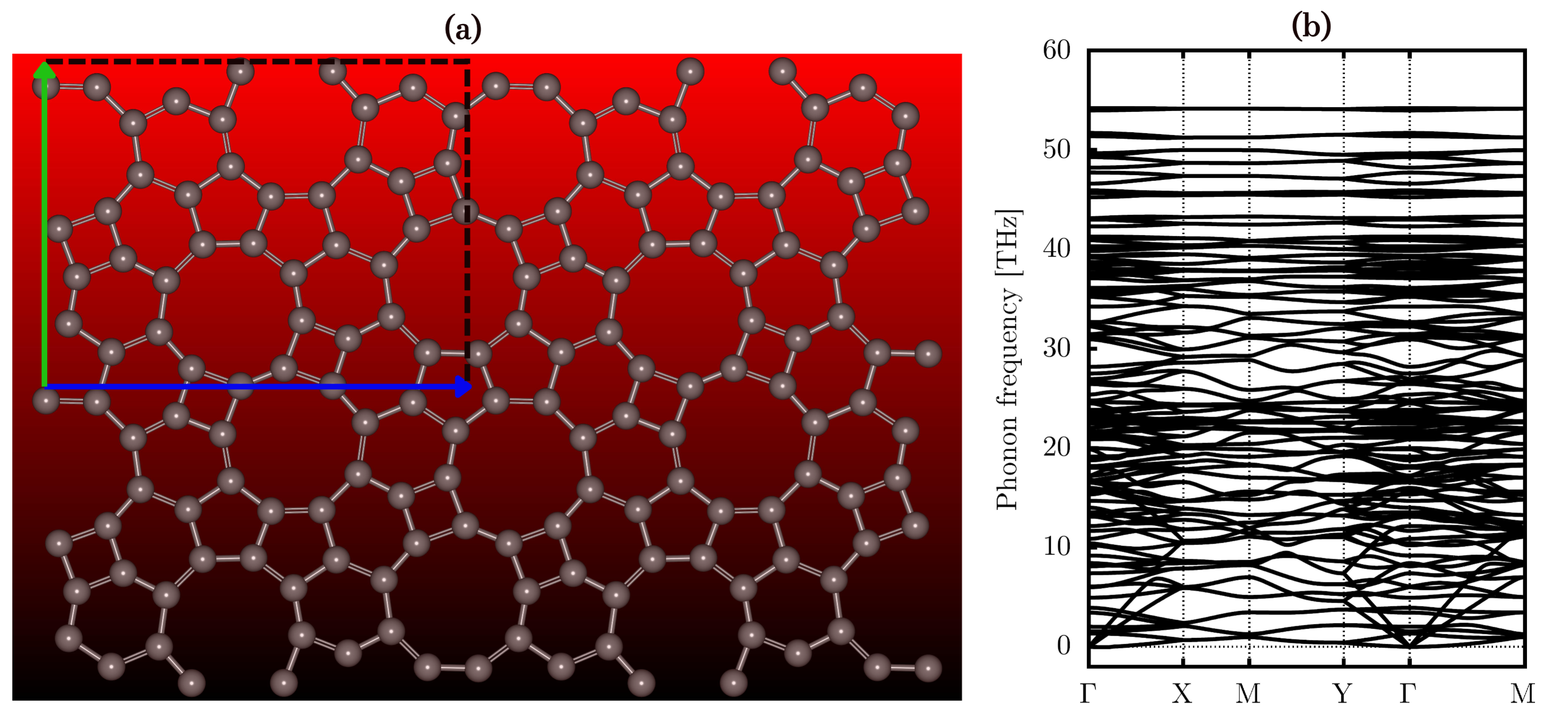}
\caption{(a) Top view of the Ennea-Graphene monolayer highlighting the rectangular unit cell (dashed black lines) and (b) phonon dispersion spectrum of its pristine lattice along high-symmetry paths in the Brillouin zone.}
\label{fig:structure_phonon}
\end{figure}

Fig.~\ref{fig:structure_phonon}(a) shows the top view of the optimized Ennea-Graphene monolayer, highlighting the rectangular unit cell (dashed black lines). The lattice is characterized by a unique combination of 4-, 5-, 6-, and 9-membered carbon rings, forming a porous and intricate framework that is distinct from the already synthesized 2D carbon materials, such as graphene,\cite{kag2021strain, morse2021hydrogenated} graphyne,\cite{liu2014hydrogen, guo2013comparative, zhang2025cli3} monolayer amorphous carbon,\cite{Toh2020} biphenylene network,\cite{Fan2021}, holey-graphyne,\cite{Liu2022} monolayer C$_{60}$ network,\cite{Hou2022,Peng2022} and graphdiyne.\cite{Jia2017} The optimized lattice parameters are found to be $a = 12.83$~\si{\angstrom} and $b = 8.99$~\si{\angstrom} corresponding to the orthorhombic crystal system, with a P2/m (No. 10) layer group. The results show a cohesive energy of \SI{-7.45}{eV/atom}, indicating energetic stability and with a value comparable to other theoretically proposed 2D carbon allotropes.\cite{Lima2025,PereiraJnior2023} 

To assess the dynamical stability of Ennea-Graphene, the phonon dispersion spectrum was computed along high-symmetry paths in the Brillouin zone (see Fig.~\ref{fig:structure_phonon}(b)). The absence of imaginary frequencies confirms the dynamical stability of the monolayer. The acoustic modes are located below \SI{5}{THz}, while the maximum optical phonon frequency reaches \SI{54}{THz} (\SI{1801}{\per\cm}), indicative of strong covalent bonding within the carbon network. The absence of a phonon band gap between acoustic and optical branches suggests significant phonon-phonon scattering, which could influence thermal transport properties.

\begin{figure}[ht]
\centering
\includegraphics[width=\textwidth]{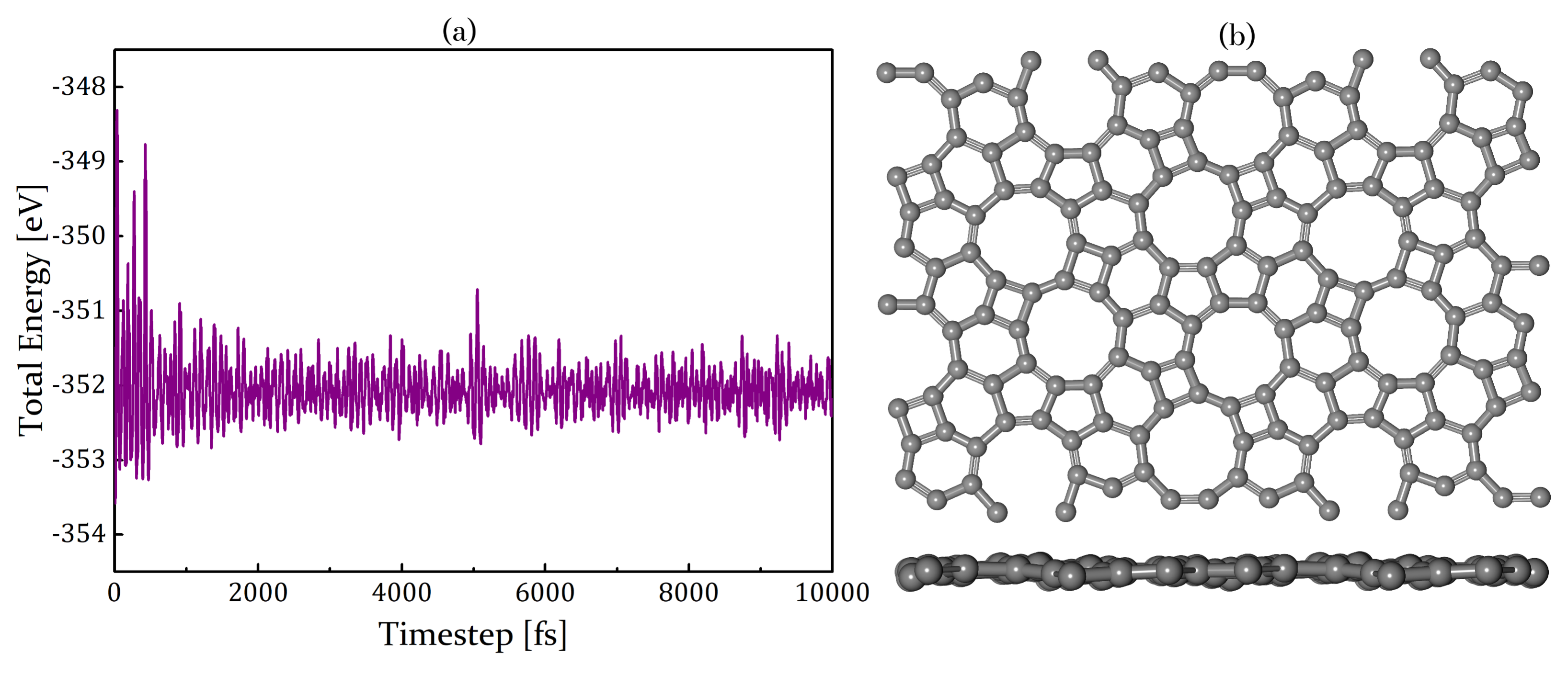}
\caption{AIMD simulation for pristine Ennea-Graphene at \SI{300}{\kelvin}. (a) Time evolution of the total energy over \SI{10000}{\fs} (\SI{10}{\ps}) of simulation. (b) Top and side views of the final structure.}
\label{fig:aimd}
\end{figure}

The thermal stability of Ennea-Graphene at room temperature was examined through AIMD simulations at \SI{300}{\kelvin}. Fig.~\ref{fig:aimd}(a) shows the time evolution of the total energy over a \SI{10}{\ps} of simulation. The system displays only minor total energy fluctuations around an average value of approximately \SI{-352}{\electronvolt}, indicating thermal equilibrium and the absence of significant structural rearrangements or phase transitions during the simulation. 

Fig.~\ref{fig:aimd})(b) presents the final configuration of the monolayer, illustrating both top and side views. The structure retains its characteristic 4-, 5-, 6-, and 9-membered ring topology throughout the simulation. The results confirm that Ennea-Graphene is thermally stable at \SI{300}{\kelvin}, further supporting its suitability for practical applications. 

\subsection{Mechanical and Electronic Properties of Ennea-Graphene}

The mechanical properties of Ennea-Graphene were evaluated by calculating its elastic constants, Young's modulus, shear modulus, and Poisson's ratio. For a 2D rectangular lattice, the Born--Huang stability criteria require $C_{11} > 0$, $C_{22} > 0$, $C_{66} > 0$, and $C_{11}C_{22} - C_{12}^{2} > 0$ \cite{Mouhat2014}. The calculated elastic constants, $C_{11} = 261.75$~\si{N/m}, $C_{22} = 264.66$~\si{N/m}, $C_{12} = 87.74$~\si{N/m}, and $C_{66} = 100.05$~\si{N/m} (Table~\ref{tab:elastic}), satisfy these conditions, confirming the mechanical stability of the monolayer.

\begin{figure}[ht]
\centering
\includegraphics[width=\textwidth]{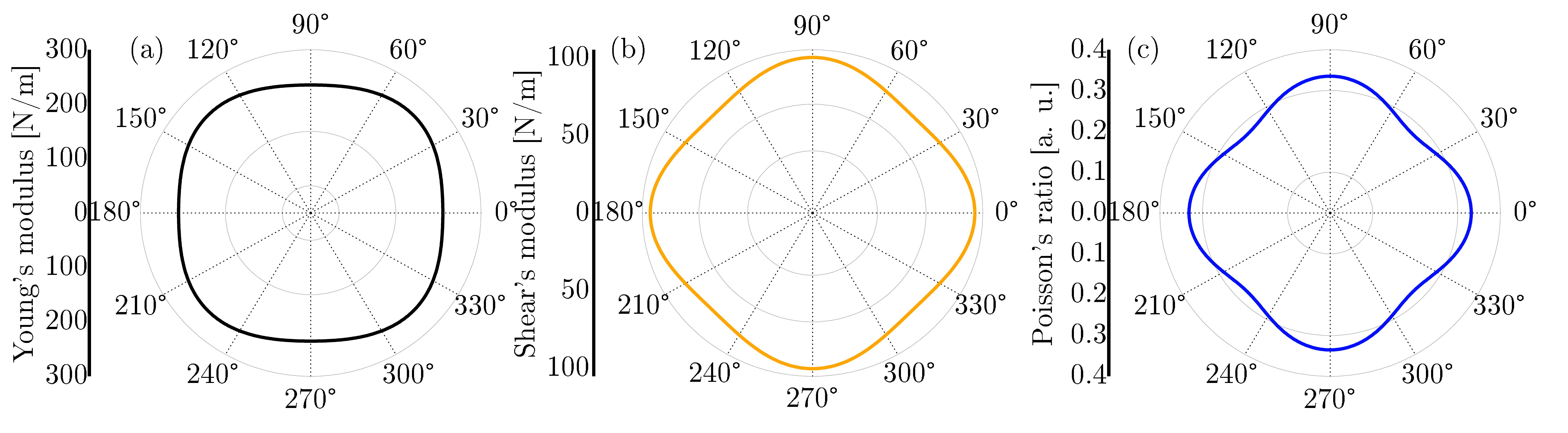}
\caption{Directional dependence of (a) Young's modulus, (b) shear modulus, and (c) Poisson's ratio for Ennea-Graphene.}
\label{fig:mech}
\end{figure}

Fig.~\ref{fig:mech} illustrates the directional dependence of the Young's modulus, shear modulus, and Poisson's ratio. The Young's modulus ranges from \SIrange{232.65}{254.89}{N/m}, showing a low anisotropy ratio of 1.10, which indicates near-isotropic in-plane stiffness. The shear modulus varies between \SIrange{87.73}{100.05}{N/m}, while Poisson's ratio lies between 0.27 and 0.34. 

\begin{table}[h!]
\begin{center}
\caption{Elastic constants $C_{ij}$ (\si{N/m}) and maximum/minimum values for Young's modulus (\si{N/m}), shear modulus (\si{N/m}), and Poisson’s ratio ($\nu$).}
\vspace{0.2 cm}
\begin{tabular}{c c c c c c c c}
\hline
Structure & $C_{11}$ & $C_{22}$ & $C_{12}$ & $C_{66}$ & $Y_{\max}/Y_{\min}$ & $G_{\max}/G_{\min}$ & $\nu_{\max}/\nu_{\min}$ \\ \hline 
Ennea-Graphene & 261.75 & 264.66 & 87.74 & 100.05 & 254.89/232.65 & 100.05/87.73 & 0.34/0.27 \\
Anisotropy ratio & -- & -- & -- & -- & 1.10 & 1.14 & 1.26 \\ \hline
\end{tabular}
\label{tab:elastic}
\end{center}
\end{table}

These elastic properties indicate that Ennea-Graphene combines mechanical robustness with a slight degree of anisotropy, which can be advantageous for specific applications where directional mechanical responses are desirable. The Young's modulus values, comparable to those of other 2D carbon allotropes such as graphdiyne \cite{Polyakova2024} ($Y_{\max}$ = 111 N/m) and biphenylene network,\cite{Luo2021} ($Y_{\max}$ = 259.7 N/m), ensure structural integrity under in-plane tensile stress. The modest anisotropy ratios of 1.10 for Young's modulus and 1.14 for shear modulus suggest that the material exhibits nearly isotropic elastic behavior. Moreover, the Poisson's ratio values, ranging from 0.27 to 0.34, fall within the typical range reported for other 2D carbon-based materials,\cite{Cheng2025, Martins2025}, implying a balanced response between lateral contraction and axial elongation. Such mechanical characteristics are crucial for ensuring durability and flexibility in potential hydrogen storage devices or composite systems incorporating Ennea-Graphene.

\begin{figure}[ht]
\centering
\includegraphics[width=0.6\textwidth]{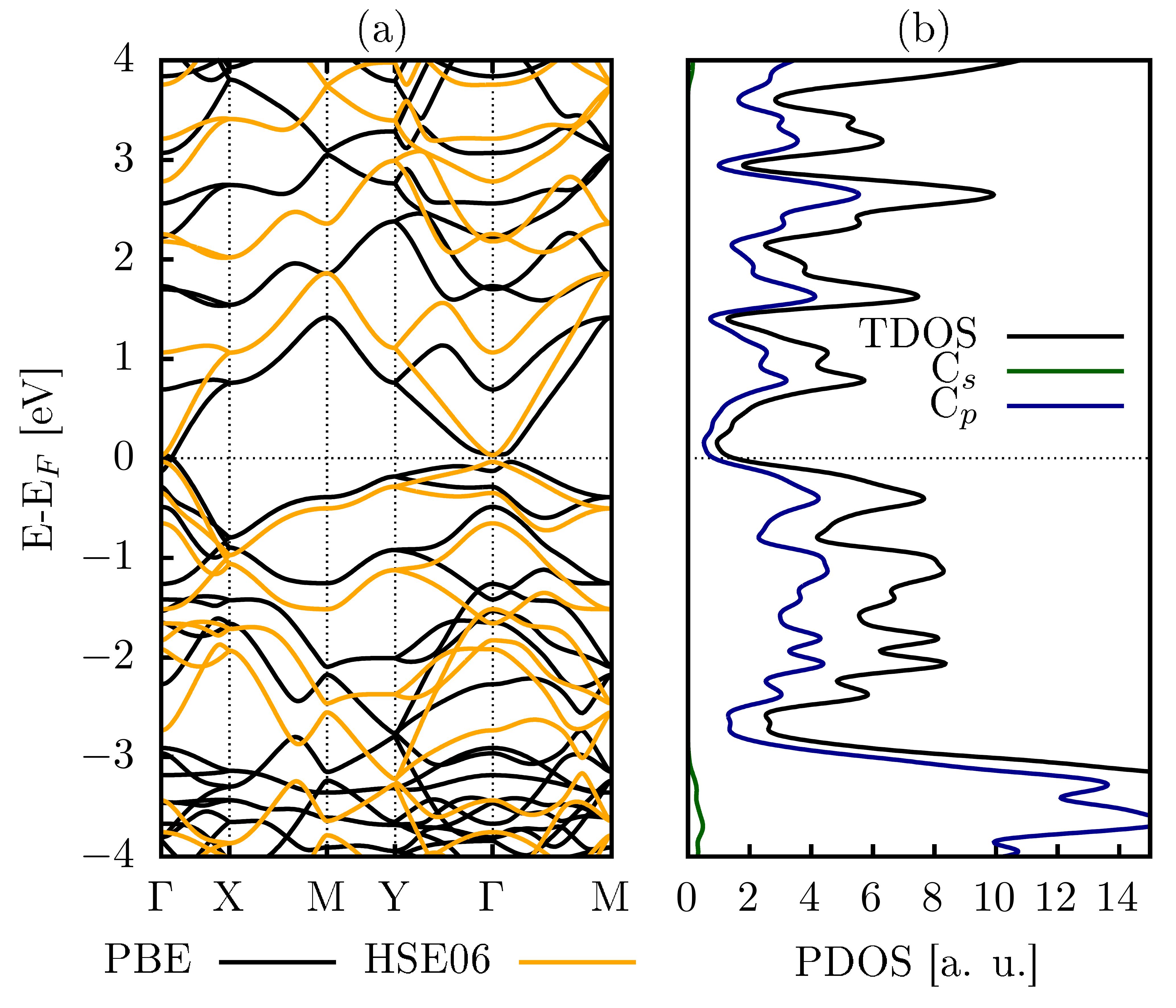}
\caption{(a) Band structure of Ennea-Graphene calculated with PBE (black) and HSE06 (orange) functionals. (b) Total and projected density of states showing contributions from carbon $s$ and $p$ orbitals.}
\label{fig:electronic}
\end{figure}

The electronic properties of Ennea-Graphene were analyzed by calculating its band structure and projected density of states (PDOS), as shown in Fig.~\ref{fig:electronic}. The band structure computed with the PBE functional reveals a metallic character, characterized by a negligible band opening of approximately \SI{0.011}{\electronvolt} near the Fermi level. To account for the known underestimation of band gaps by the PBE functional, the HSE06 hybrid functional was employed, which yields a slightly larger opening of \SI{0.067}{\electronvolt}. However, this small value remains within the expected range of computational uncertainty, supporting the classification of Ennea-Graphene as a material with metallic-like properties.

The PDOS analysis indicates that carbon $p$ orbitals dominate the valence and conduction band edges, while $s$ orbitals contribute only marginally. This orbital character is consistent with the $\pi$-conjugated nature of the carbon framework. The metallic-like electronic structure, combined with the mechanical robustness of Ennea-Graphene, suggests that the material can sustain charge transfer interactions essential for the adsorption of sodium atoms and hydrogen molecules, which are crucial for subsequent storage applications.

\subsection{Sodium Decoration on Ennea-Graphene Monolayer}

\begin{figure}[ht]
\centering
\includegraphics[width=0.7\textwidth]{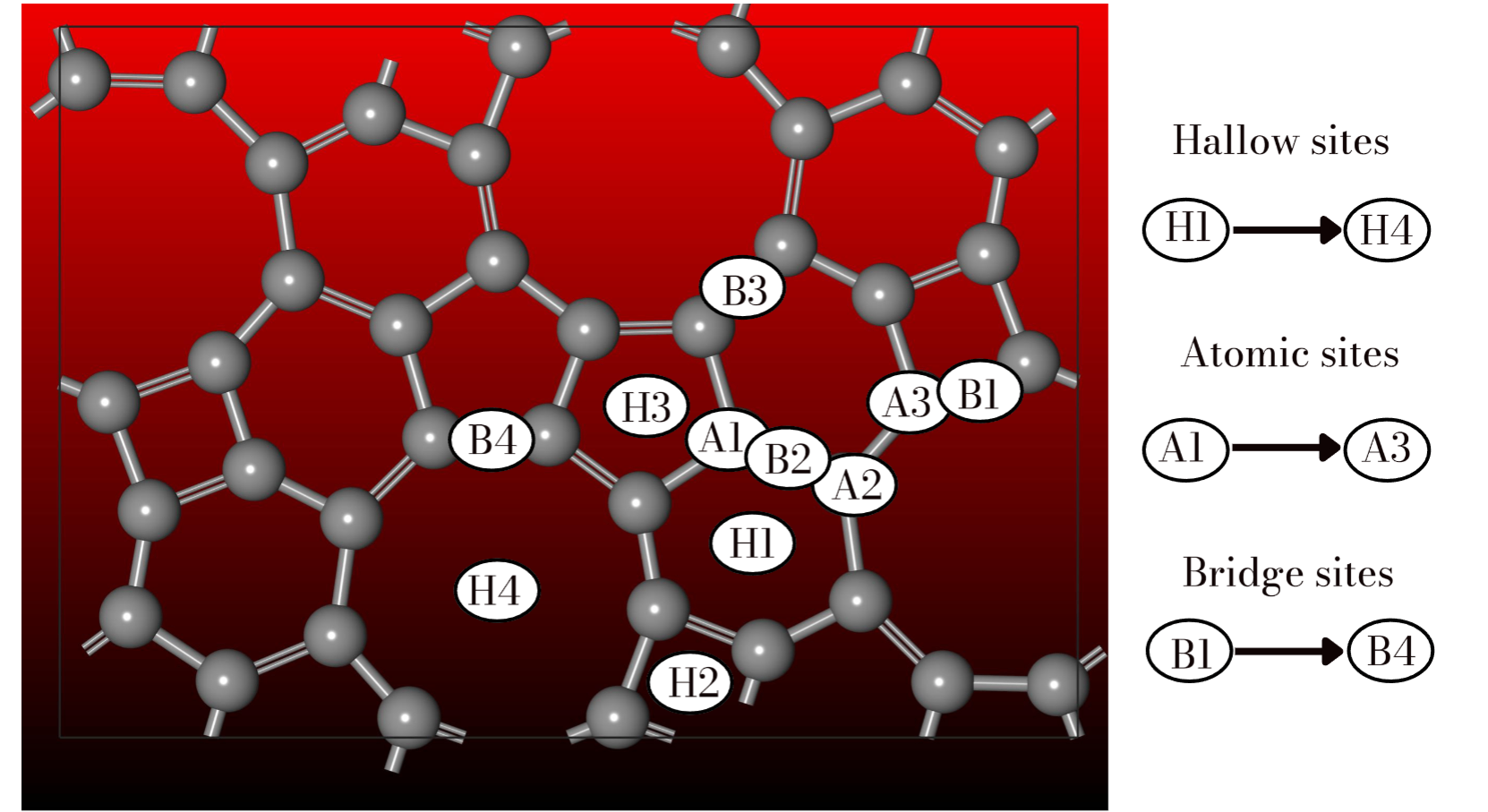}
\caption{Distinct sodium adsorption sites on Ennea-Graphene: hollow (H1--H4), atomic (A1--A3), and bridge (B1--B4) positions.}
\label{fig:na_sites}
\end{figure}

To explore the preferential binding sites for sodium atoms on the Ennea-Graphene monolayer, single \ce{Na} atoms were placed at a range of high-symmetry positions and the resultant structures were relaxed in terms of their internal coordinates. As depicted in Fig.~\ref{fig:na_sites}, the tested sites include hollow positions (H1 to H4), on-top atomic positions (A1 to A3), and bridge positions (B1 to B4), which collectively capture the diverse local environments within the mixed 4-, 5-, 6-, and 9-membered ring lattice.

Adsorption energies were calculated using the expression
\begin{equation}
E_{\text{ads}} = E_{\text{Na@Ennea}} - E_{\text{Ennea}} - E_{\text{Na}},   
\end{equation}
\noindent where $E_{\text{Na@Ennea}}$ is the total energy of the sodium-decorated system, $E_{\text{Ennea}}$ refers to the pristine monolayer, and $E_{\text{Na}}$ is the energy of an isolated sodium atom. Results highlight that sodium atoms preferentially bind at the centers of the nonagonal rings (H4 sites), where they exhibit the strongest adsorption energies. This preference underscores the significance of the rare nonagonal motifs in stabilizing sodium adsorption. 

\begin{figure}[ht]
\centering
\includegraphics[width=\textwidth]{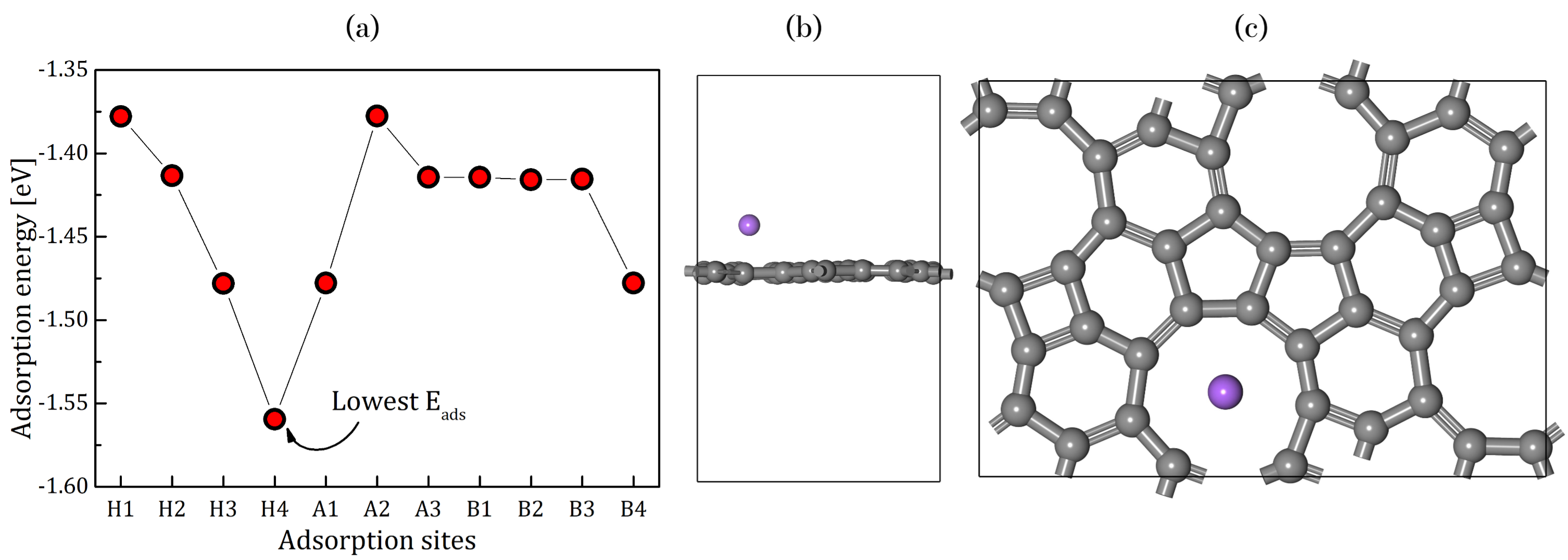}
\caption{(a) Adsorption energies for single \ce{Na} atoms on each available site of Ennea-Graphene. (b) Side and (c) top view of the most stable configuration at the H4 site.}
\label{fig:na_adsorption}
\end{figure}

Fig.~\ref{fig:na_adsorption} summarizes the adsorption energy results for single sodium atoms on the various investigated sites. The H4 site, located at the center of the nonagonal ring, exhibits the most negative adsorption energy (\SI{-1.56}{eV}), confirming it as the most favorable binding site (see Fig.~\ref{fig:na_adsorption}(a)). Other hollow, bridge, and atomic sites show slightly less negative adsorption energies, with values ranging from \SIrange{-1.38}{-1.49}{\electronvolt}. These findings reinforce the role of the nonagonal rings as preferential adsorption centers in Ennea-Graphene, offering an energetically favorable environment for sodium binding.

The structural configuration of the most stable adsorption geometry is illustrated in Figs.~\ref{fig:na_adsorption}(b) and~\ref{fig:na_adsorption}(c). Sodium atoms remain well anchored above the monolayer without inducing significant distortions in the carbon lattice. This stable adsorption is crucial for the subsequent reversible hydrogen storage process. The inclusion of van der Waals interactions throughout these calculations ensures a reliable description of the binding energetics and geometry. All computed adsorption energies for the investigated sites are compiled in Table~\ref{Table-Eads}.

\begin{table}[h!]
\centering
\caption{Adsorption energies ($E_{\text{ads}}$) for the adsorption sites evaluated during Na decoration on Ennea-Graphene.}
\vspace{0.2 cm}
\resizebox{\textwidth}{!}{%
\begin{tabular}{c c c c c c c c c c c c}
\hline
 structure site & H1 & H2 & H3 & H4 & A1 & A2 & A3 & B1 & B2 & B3 & B4\\ \hline 
 \ce{E_{ads}} (eV) & \num{-1.478} & \num{-1.413} & \num{-1.478} & \num{-1.559} & \num{-1.478} & \num{-1.378} & \num{-1.414} & \num{-1.414} & \num{-1.416}  & \num{-1.416} & \num{-1.478}  \\ \hline
\end{tabular}%
}
\label{Table-Eads}
\end{table}

\begin{figure}[ht]
\centering
\includegraphics[width=0.9\textwidth]{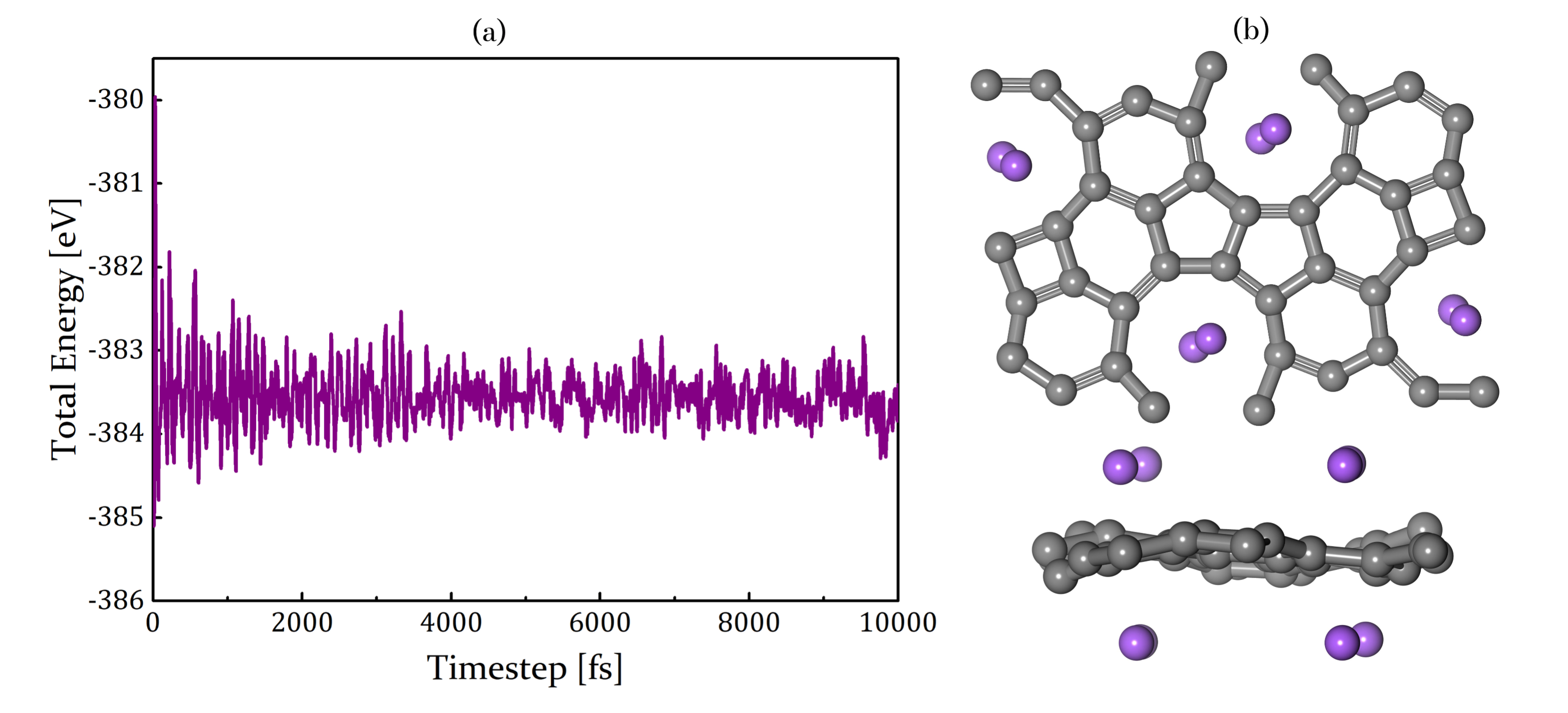}
\caption{AIMD simulation for Ennea-Graphene with 8 \ce{Na} atoms at \SI{300}{\kelvin}. (a) Time evolution of the total energy over \SI{10000}{\fs} of simulation. (b) Top and side views of the final structure.}
\label{fig:aimd_na}
\end{figure}

To verify the thermal stability of the sodium-decorated structure, AIMD simulations at \SI{300}{\kelvin} were conducted with eight sodium atoms placed at the most favorable H4 sites. Fig.~\ref{fig:aimd_na}(a) presents the total energy evolution over \SI{10}{\ps} of simulation. The system exhibits only small fluctuations around an average value of approximately \SI{-384}{\electronvolt}, suggesting that no significant desorption, structural reconstruction, or phase transition occurs during the simulation.

Final configurations, depicted in Fig.~\ref{fig:aimd_na}(b), show that the sodium atoms remain firmly anchored at the nonagonal ring centers with minimal displacement. The carbon framework also retains its integrity, preserving the mixed-ring topology characteristic of Ennea-Graphene. These results provide strong evidence for the thermal stability of the Na-decorated structure under ambient conditions, supporting its viability for hydrogen storage applications.

\begin{figure}[ht]
\centering
\includegraphics[width=0.6\textwidth]{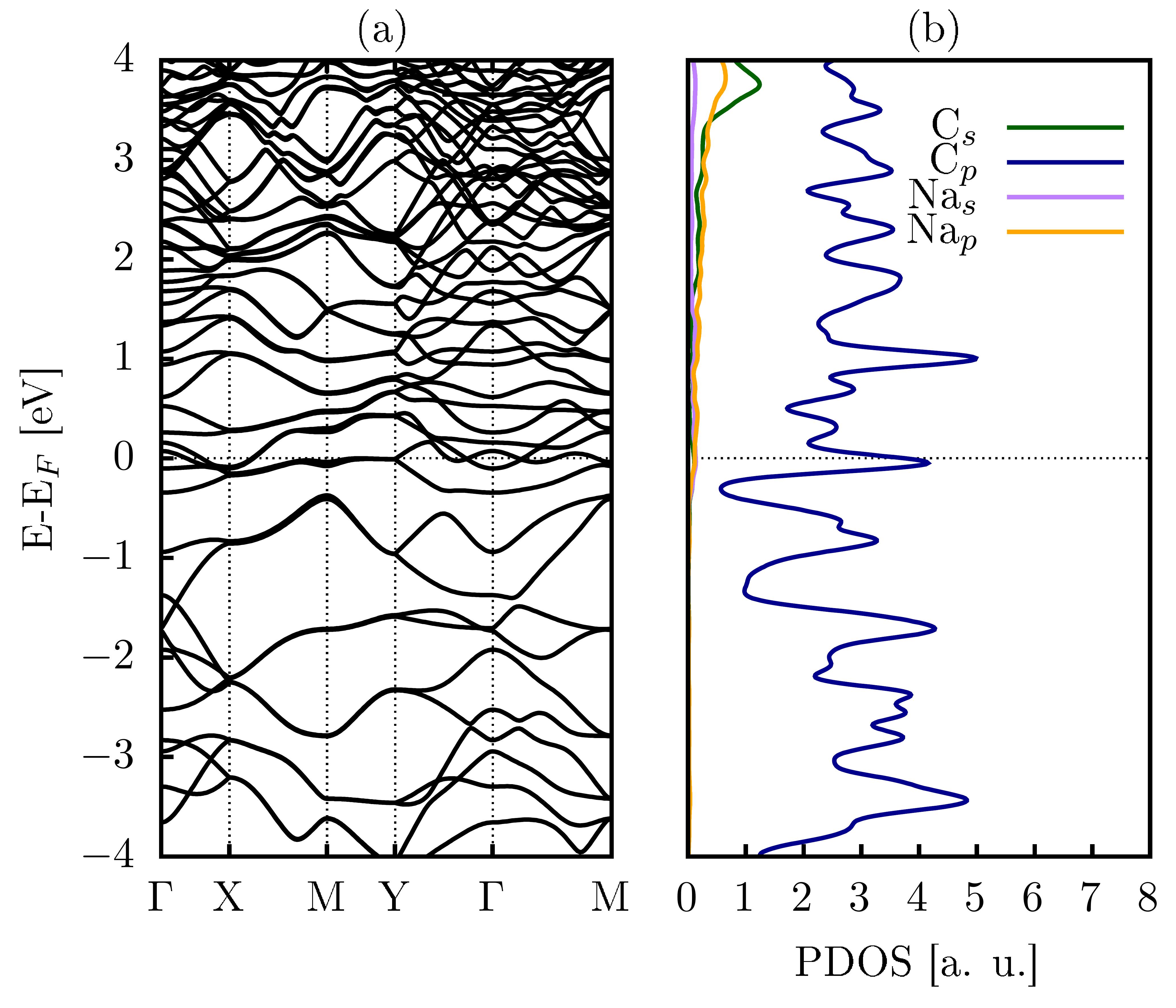}
\caption{(a) Band structure and (b) projected density of states for Ennea-Graphene with 8 \ce{Na} atoms. Contributions from carbon $s$ and $p$ orbitals and sodium $s$ and $p$ orbitals are indicated.}
\label{fig:na_electronic}
\end{figure}

The electronic structure of the Ennea-Graphene monolayer decorated with eight sodium atoms was further examined through band structure and density of states analyses. As displayed in Fig.~\ref{fig:na_electronic}(a), the system still exhibits metallic behavior, with bands crossing the Fermi level along all high-symmetry directions of the Brillouin zone. This metallicity marks a clear transition from the tiny gap opening in the pristine structure to a fully conductive state upon sodium decoration.

PDOS, shown in Fig.~\ref{fig:na_electronic}(b), reveals that carbon $p$ orbitals dominate the electronic states near the Fermi level, with contributions from sodium $s$ and $p$ orbitals. These sodium states hybridize with the carbon framework, enhancing the density of conduction states and favoring the charge transfer processes that are critical for efficient hydrogen adsorption. The metallic nature of the \ce{Na}-decorated system is thus consistent with its suitability as a platform for reversible hydrogen storage.

\subsection{Hydrogen Storage Properties of Ennea-Graphene}

\begin{figure}[ht]
\centering
\includegraphics[width=0.7\textwidth]{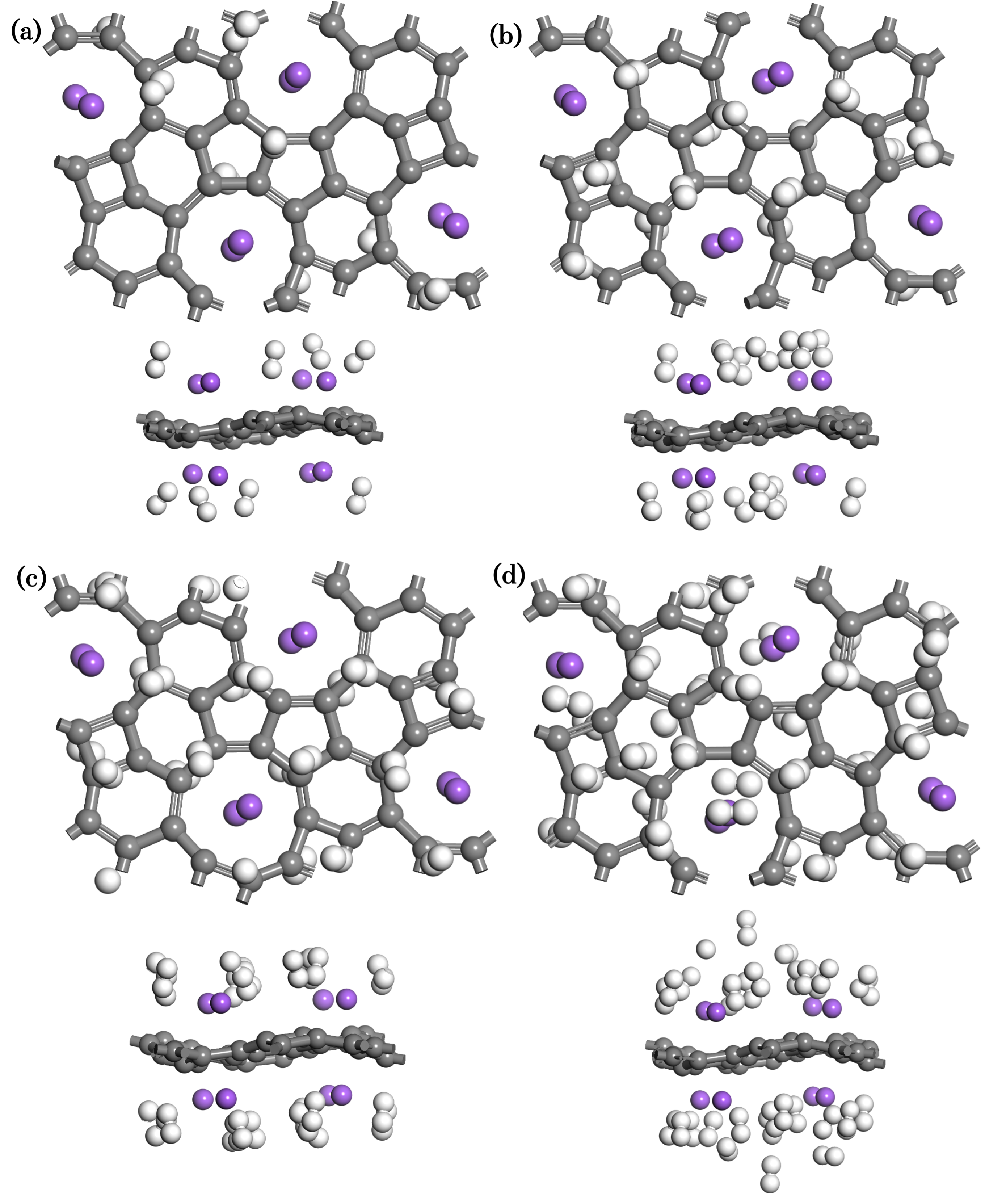}
\caption{\ce{H2} gradual saturation behavior on Na@Ennea-Graphene: (a) 8 \ce{H2}, (b) 16 \ce{H2}, (c) 24 \ce{H2}, and (d) 32 \ce{H2} configurations.}
\label{fig:h2_structures}
\end{figure}

The hydrogen storage potential of the \ce{Na}-decorated Ennea-Graphene (Na@Ennea) system was evaluated by sequentially adsorbing one \ce{H2} molecule per sodium atom, forming configurations containing 8, 16, 24, and 32 \ce{H2} molecules per unit cell. This approach resulted in a fully saturated configuration, accommodating 32 \ce{H2} molecules, corresponding to four hydrogen molecules per sodium atom. Fig.~\ref{fig:h2_structures} illustrates the progressive adsorption and saturation behavior.

To quantify the adsorption performance, the average adsorption energy ($E_{\text{ads}}$) was computed, the consecutive adsorption energy ($E_{\text{con}}$), hydrogen adsorption capacity (HAC), and estimated desorption temperature ($T_{\text{des}}$). The average adsorption energy was calculated as:

\begin{equation}
E_{\text{ads}} = \frac{1}{n} \left( E_{\text{Na@Ennea}+n\text{H}_2} - E_{\text{Na@Ennea}} - n E_{\text{H}_2} \right),
\end{equation}

\noindent where $E_{\text{Na@Ennea}+n\text{H}_2}$ represents the total energy of the \ce{Na}-decorated Ennea-Graphene with $n$ adsorbed \ce{H2} molecules, $E_{\text{Na@Ennea}}$ is the energy of the \ce{Na}-decorated monolayer, and $E_{\text{H}_2}$ corresponds to the energy of an isolated \ce{H2} molecule.

Hydrogen adsorption capacity was expressed in weight percentage:
\begin{equation}
\text{HAC (wt\%)} = \frac{n_{\text{H}} M_{\text{H}}}{n_{\text{C}} M_{\text{C}} + n_{\text{Na}} M_{\text{Na}} + n_{\text{H}} M_{\text{H}}},
\end{equation}
\noindent where $n_{\text{C}}$, $n_{\text{Na}}$, and $n_{\text{H}}$ denote the number of carbon, sodium, and hydrogen atoms, respectively, and $M_{\text{C}}$, $M_{\text{Na}}$, and $M_{\text{H}}$ are their molar masses. The hydrogen adsorption capacity increased linearly with coverage, ranging from 2.35~wt\si{\percent} for Na@Ennea + 8\ce{H2} to 8.80~wt\si{\percent} at full saturation (32\ce{H2}). 

The average adsorption energies (Table~\ref{Table-Na@FFSN-graphene}) range between \SI{-0.18}{\electronvolt} and \SI{-0.15}{\electronvolt}, supporting the physisorption nature of the interaction, as further confirmed by the nearly unchanged \ce{H}--\ce{H} bond length of $\sim 0.76$~\si{\angstrom}, close to the \SI{0.75}{\angstrom} of a free \ce{H2} molecule.
On the other hand, assuming atmospheric pressure (\SI{1}{atm}), the hydrogen desorption temperature (\ce{T_{des}}) was estimated using the van’t Hoff equation \cite{Alhameedi2019,Kanmani2014}:

\begin{equation}
T_{\text{des}} = \frac{\lvert E_{\text{ads}} \rvert \, R}{k_{B} \, \Delta S}
\end{equation}

Here, R denotes the universal gas constant, $k_{B}$ is the Boltzmann constant, and $\Delta S$ corresponds to the entropy change associated with the hydrogen phase transition from gas to liquid (\SI{75.44}{\joule\per\mol\kelvin}). The results show that the desorption temperature remains within a practical range from \SIrange{197}{232}{\kelvin}. The lowest \ce{T_{des}} observed correspond to 32\ce{H2} Na@Ennea-Graphene, while the highest  \ce{T_{des}} is for 24\ce{H2} Na@Ennea-Graphene.

\begin{table}[h!]
\begin{center}
\caption{Adsorption energy ($E_{\text{ads}}$), hydrogen adsorption capacity (HAC), average \ce{H}--\ce{H} bond length ($R_{\text{H--H}}$), and desorption temperature (\ce{T_{des}}) for Na@Ennea-Graphene + n\ce{H2}.}
\vspace{0.2 cm}
\begin{tabular}{c c c c c c}
\hline
 system &  \ce{E_{ads}}(\si{\electronvolt})  & HAC(wt\si{\percent}) & \ce{R_{H-H}} ( \si{\angstrom}) & \ce{T_{des}(\si{\kelvin})}\\ \hline 
 Na@Ennea-Graphene + 8\ce{H2}  & \num{-0.168} & 2.35 & 0.76 & 215\\
 Na@Ennea-Graphene + 16\ce{H2} & \num{-0.176} & 4.60 & 0.76 & 224\\
 Na@Ennea-Graphene + 24\ce{H2} & \num{-0.181} & 6.75 & 0.76 & 232\\
 Na@Ennea-Graphene + 32\ce{H2} & \num{-0.154} & 8.80 & 0.76 & 197 \\
 \hline
\end{tabular}
\label{Table-Na@FFSN-graphene}
\end{center}
\end{table}

\begin{figure}[ht]
\centering
\includegraphics[width=0.9\textwidth]{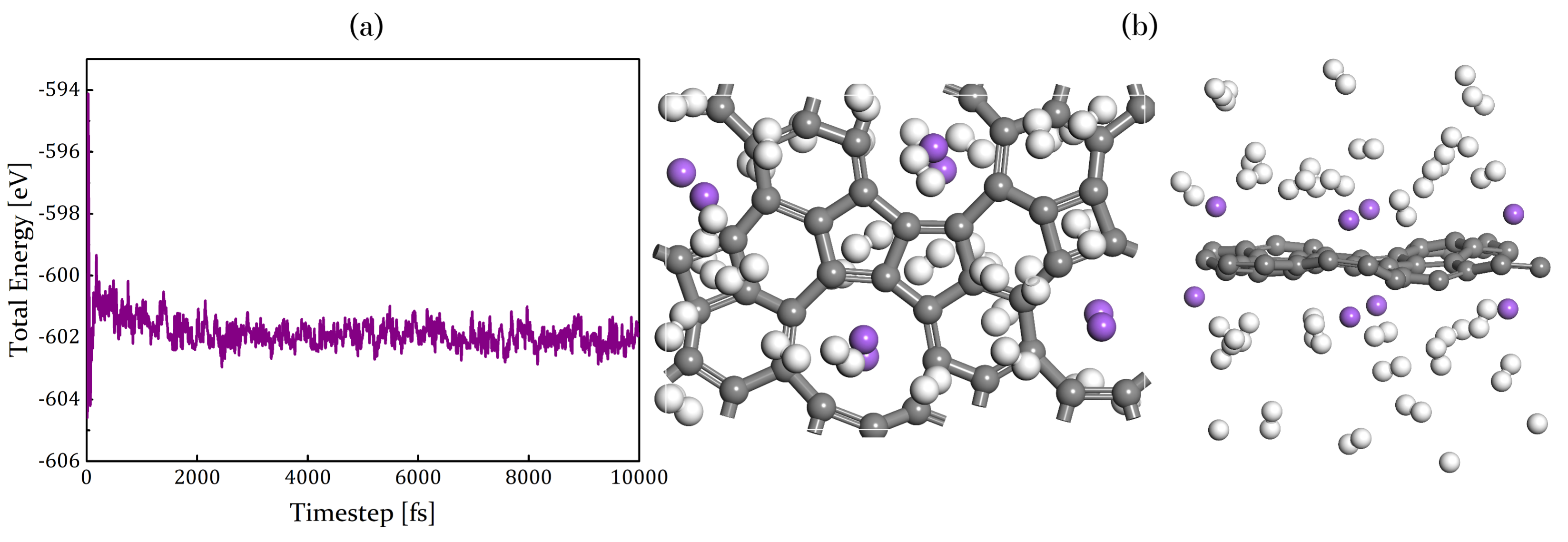}
\caption{AIMD simulation results for Na@Ennea-Graphene + 32 \ce{H2} at \SI{300}{\kelvin}. (a) Total energy evolution over \SI{10000}{\fs} (\SI{10}{\ps}). (b) Top and side views of the final structure after simulation.}
\label{fig:aimd_h2}
\end{figure}

To assess the thermal stability of Na@Ennea-Graphene in the presence of hydrogen, AIMD simulations were conducted at \SI{300}{\kelvin} for the Na@Ennea + 32 \ce{H2} system. The simulations ran for \SI{10}{\ps}, with the total energy evolution shown in Fig.~\ref{fig:aimd_h2}(a). The energy profile exhibits abrupt fluctuations that correspond to the desorption of \ce{H2} molecules, providing direct evidence of reversible hydrogen storage and thermally activated release under near-ambient conditions.

Despite the desorption events, the carbon substrate retains its structural integrity throughout the simulation. No significant atomic rearrangement or bond dissociation occurs, and the sodium atoms remain anchored at their preferred nonagonal ring adsorption sites, as seen in Fig.~\ref{fig:aimd_h2}(b). Only minor displacements of the \ce{Na} atoms are observed, further supporting the stability of the decorated structure.

These results demonstrate that Na@Ennea-Graphene not only provides high hydrogen storage capacity but also enables the reversible release of hydrogen without compromising structural stability. This behavior is a key requirement for practical hydrogen storage materials. Furthermore, the preservation of the monolayer’s framework during desorption suggests that the material could withstand multiple adsorption–desorption cycles, a critical factor for real-world applications in hydrogen energy systems.    

\begin{figure}[ht]
\centering
\includegraphics[width=0.9\textwidth]{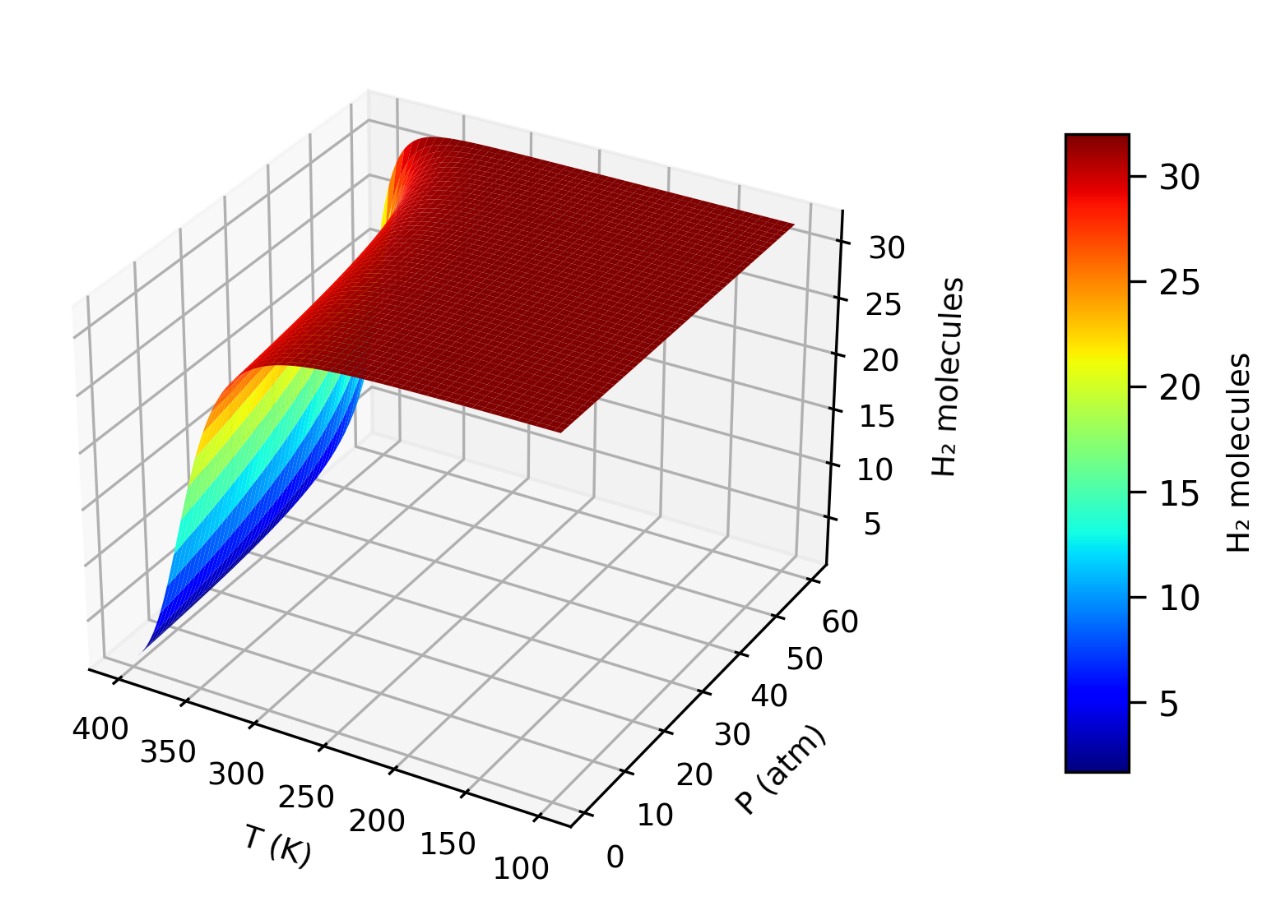}
\caption{The average number of absorbed \ce{H2} on Na@Ennea-Graphene at various pressures P(\si{atm}) and temperatures T(\si{\kelvin}).}
\label{fig:H2-T-P}
\end{figure}

In Fig.~\ref{fig:H2-T-P}, the thermodynamic response of \ce{H2} uptake and release on Na@Ennea-Graphene is shown. The diagram incorporates both entropic and enthalpic contributions, allowing a more reliable evaluation of the number of adsorbed hydrogen molecules ($n$) as a function of pressure ($P$) and temperature ($T$). The selected operational conditions are $P = 30/3$~\si{atm} and $T = 25^\circ$C/$100^\circ$\si{C} for adsorption and desorption, respectively. Our analysis reveals that at $P = 30$~\si{atm} and $T = 25^\circ$\si{C}, approximately 31.24 \ce{H2} molecules are retained, while at $P = 3$~\si{atm} and $T = 100^\circ$\si{C}, adsorption is nearly suppressed. This leads to a reversible storage capacity of 32 \ce{H2} molecules, corresponding to 8.80~wt\si{\percent}. These results demonstrate that Na@Ennea-Graphene represents an efficient platform for hydrogen storage under realistic conditions.

\section{Conclusion}

In this study, a novel two-dimensional carbon allotrope, Ennea-Graphene, was proposed and systematically investigated via DFT simulations. This novel lattice is distinguished by its combination of 4-, 5-, 6-, and 9-membered carbon rings. DFT calculations demonstrated that the pristine structure is mechanically, thermally, and dynamically stable, with a near-isotropic elastic response and metallic-like electronic properties. Sodium atoms preferentially adsorb at the centers of the nonagonal rings with strong binding energies, and the resulting \ce{Na}-decorated system preserves its structural integrity under thermal fluctuations at room temperature. Hydrogen storage performance was evaluated through a series of adsorption configurations, reaching a maximum of four \ce{H2} molecules per sodium atom and achieving a gravimetric capacity of 8.80 wt\si{\percent}. The adsorption energies and \ce{H}--\ce{H} bond lengths indicated that hydrogen molecules are physisorbed, enabling reversible storage. AIMD simulations further confirmed that hydrogen desorption can occur under near-ambient conditions without compromising the stability of the Na-decorated monolayer.

These findings establish \ce{Na}-decorated Ennea-Graphene as a promising material for high-capacity, reversible hydrogen storage applications. The unique nonagonal-ring motifs and their role in stabilizing sodium adsorption open new avenues for the design of advanced carbon-based hydrogen storage systems. 

\section*{Acknowledgment}
B. D. A. H. acknowledges the support of CAPES, a Brazilian funding agency, for the PhD scholarship. The authors also express their gratitude to the National Laboratory for Scientific Computing for providing resources through the Santos Dumont supercomputer, and to the ``Centro Nacional de Processamento de Alto Desempenho em S\~ao Paulo'' (CENAPAD-SP, UNICAMP/FINEP - MCTI project) for support related to projects 897 and Lobo Carneiro HPC (project 133). Also, this work was supported by the Brazilian funding agencies Fundação de Amparo à Pesquisa do Estado de São Paulo - FAPESP (grant no. 2022/03959-6, 2022/00349- 2, 2022/14576-0, 2020/01144-0, 2024/05087-1, and 2022/16509-9), and National Council for Scientific and Technological Development - CNPq (grant no. 307213/2021–8). L.A.R.J. acknowledges the financial support from FAP-DF grants 00193.00001808/2022-71 and $00193-00001857/2023-95$, FAPDF-PRONEM grant 00193.00001247/2021-20, PDPG-FAPDF-CAPES Centro-Oeste 00193-00000867/2024-94, and CNPq grants $350176/2022-1$ and $167745/2023-9$. A.C.D acknowledges the financial support from FAP-DF grants 00193-00001817/2023-43 and 00193-00002073/2023-84, and from CNPq grants 408144/2022-0, 305174/2023-1, 444069/2024-0 and 444431/2024-1.

\printcredits
\bibliographystyle{elsarticle-num}
\bibliography{references}

\begin{thebibliography}{10}
\expandafter\ifx\csname url\endcsname\relax
  \def\url#1{\texttt{#1}}\fi
\expandafter\ifx\csname urlprefix\endcsname\relax\def\urlprefix{URL }\fi
\expandafter\ifx\csname href\endcsname\relax
  \def\href#1#2{#2} \def\path#1{#1}\fi

\bibitem{guilbert2021hydrogen}
D.~Guilbert, G.~Vitale, Hydrogen as a clean and sustainable energy vector for global transition from fossil-based to zero-carbon, Clean Technologies 3~(4) (2021) 881--909.

\bibitem{cheekatamarla2024hydrogen}
P.~Cheekatamarla, Hydrogen and the global energy transition—path to sustainability and adoption across all economic sectors, Energies 17~(4) (2024) 807.

\bibitem{sadeq2024hydrogen}
A.~M. Sadeq, R.~Z. Homod, A.~K. Hussein, H.~Togun, A.~Mahmoodi, H.~F. Isleem, A.~R. Patil, A.~H. Moghaddam, Hydrogen energy systems: Technologies, trends, and future prospects, Science of The Total Environment 939 (2024) 173622.

\bibitem{tashie2021hydrogen}
B.~C. Tashie-Lewis, S.~G. Nnabuife, Hydrogen production, distribution, storage and power conversion in a hydrogen economy-a technology review, Chemical Engineering Journal Advances 8 (2021) 100172.

\bibitem{sun2023brief}
C.~Sun, C.~Wang, T.~Ha, J.~Lee, J.-H. Shim, Y.~Kim, A brief review of characterization techniques with different length scales for hydrogen storage materials, Nano Energy 113 (2023) 108554.

\bibitem{ma2024large}
N.~Ma, W.~Zhao, W.~Wang, X.~Li, H.~Zhou, Large scale of green hydrogen storage: Opportunities and challenges, International Journal of Hydrogen Energy 50 (2024) 379--396.

\bibitem{heinemann2021enabling}
N.~Heinemann, J.~Alcalde, J.~M. Miocic, S.~J. Hangx, J.~Kallmeyer, C.~Ostertag-Henning, A.~Hassanpouryouzband, E.~M. Thaysen, G.~J. Strobel, C.~Schmidt-Hattenberger, et~al., Enabling large-scale hydrogen storage in porous media--the scientific challenges, Energy \& Environmental Science 14~(2) (2021) 853--864.

\bibitem{DOE2025HydrogenStorageTargets}
{U.S. Department of Energy}, \href{https://www.energy.gov/eere/fuelcells/doe-technical-targets-onboard-hydrogen-storage-light-duty-vehicles}{Doe technical targets for onboard hydrogen storage for light-duty vehicles} (2025).
\newline\urlprefix\url{https://www.energy.gov/eere/fuelcells/doe-technical-targets-onboard-hydrogen-storage-light-duty-vehicles}

\bibitem{CHEN2024510}
X.~Chen, J.~Li, L.~Zhang, N.~Wang, J.~Cheng, Z.~Ma, P.~Gao, G.~Wang, X.~Cai, D.~Guo, J.~Xiang, L.~Zhang, \href{https://www.sciencedirect.com/science/article/pii/S0360319924009212}{Computational evaluation of li-decorated $\alpha$-c3n2 as a room temperature reversible hydrogen storage medium}, International Journal of Hydrogen Energy 62 (2024) 510--519.
\newblock \href {https://doi.org/https://doi.org/10.1016/j.ijhydene.2024.03.089} {\path{doi:https://doi.org/10.1016/j.ijhydene.2024.03.089}}.
\newline\urlprefix\url{https://www.sciencedirect.com/science/article/pii/S0360319924009212}

\bibitem{LIU2025105802}
Z.~Liu, X.~Chen, Y.~Liao, L.~Zhang, J.~A. Laranjeira, \href{https://www.sciencedirect.com/science/article/pii/S2468023025000653}{First-principles insights of na-decorated b7n5 monolayer for advanced hydrogen storage}, Surfaces and Interfaces 58 (2025) 105802.
\newblock \href {https://doi.org/https://doi.org/10.1016/j.surfin.2025.105802} {\path{doi:https://doi.org/10.1016/j.surfin.2025.105802}}.
\newline\urlprefix\url{https://www.sciencedirect.com/science/article/pii/S2468023025000653}

\bibitem{djebablia2024metal}
I.~Djebablia, Y.~Z. Abdullahi, K.~Zanat, F.~Ersan, Metal-decorated boron phosphide (bp) biphenylene and graphenylene networks for ultrahigh hydrogen storage, International Journal of Hydrogen Energy 66 (2024) 33--39.

\bibitem{ABDULLAHI2025116631}
Y.~Z. Abdullahi, J.~A. Laranjeira, J.~R. Sambrano, \href{https://www.sciencedirect.com/science/article/pii/S2352152X25013441}{β-naphthyldiene: A novel multifunctional 2d material for energy storage applications}, Journal of Energy Storage 122 (2025) 116631.
\newblock \href {https://doi.org/https://doi.org/10.1016/j.est.2025.116631} {\path{doi:https://doi.org/10.1016/j.est.2025.116631}}.
\newline\urlprefix\url{https://www.sciencedirect.com/science/article/pii/S2352152X25013441}

\bibitem{abifarin20242d}
J.~K. Abifarin, J.~F. Torres, Y.~Lu, 2d materials for enabling hydrogen as an energy vector, Nano Energy (2024) 109997.

\bibitem{ghotia2025review}
S.~Ghotia, P.~Kumar, A.~K. Srivastava, A review on 2d materials: unveiling next-generation hydrogen storage solutions, advancements and prospects, Journal of Materials Science 60~(3) (2025) 1071--1097.

\bibitem{kag2021strain}
D.~Kag, N.~Luhadiya, N.~D. Patil, S.~Kundalwal, Strain and defect engineering of graphene for hydrogen storage via atomistic modelling, International Journal of Hydrogen Energy 46~(43) (2021) 22599--22610.

\bibitem{morse2021hydrogenated}
J.~R. Morse, D.~A. Zugell, E.~Patterson, J.~W. Baldwin, H.~D. Willauer, Hydrogenated graphene: Important material properties regarding its application for hydrogen storage, Journal of Power Sources 494 (2021) 229734.

\bibitem{ma2024li}
L.-J. Ma, Y.~Sun, J.~Jia, H.-S. Wu, Li-decorated b-doped biphenylene network for reversible hydrogen storage, Fuel 357 (2024) 129652.

\bibitem{chotsawat2024first}
M.~Chotsawat, L.~Ngamwongwan, P.~Falun, S.~Jungthawan, A.~Junkaew, S.~Suthirakun, First-principles screening of metal-decorated biphenylene as efficient hydrogen storage materials, International Journal of Hydrogen Energy 81 (2024) 573--581.

\bibitem{liu2014hydrogen}
Y.~Liu, W.~Liu, R.~Wang, L.~Hao, W.~Jiao, Hydrogen storage using na-decorated graphyne and its boron nitride analog, International journal of hydrogen energy 39~(24) (2014) 12757--12764.

\bibitem{guo2013comparative}
Y.~Guo, X.~Lan, J.~Cao, B.~Xu, Y.~Xia, J.~Yin, Z.~Liu, A comparative study of the reversible hydrogen storage behavior in several metal decorated graphyne, International journal of hydrogen energy 38~(10) (2013) 3987--3993.

\bibitem{zhang2025cli3}
Z.~Zhang, H.~Chen, Cli3-decorated $\gamma$-graphyne nanosheet for efficient hydrogen storage, International Journal of Hydrogen Energy 146 (2025) 149996.

\bibitem{jiang2023density}
Q.~Jiang, X.~Bai, Z.~Jia, S.~Lu, P.~Song, Y.~Chen, P.~Shan, H.~Cui, R.~Feng, Q.~Kang, et~al., Density functional theory study of superalkali nli4-decorated graphdiyne nanosheets as hydrogen storage materials, ACS Applied Nano Materials 6~(15) (2023) 14063--14075.

\bibitem{bajgirani2024boosting}
M.~A. Bajgirani, Z.~Biglari, M.~Sahihi, Boosting hydrogen storage capacity in modified-graphdiyne structures: A comprehensive density functional study, Materials Today Communications 39 (2024) 108787.

\bibitem{laranjeira2025tphe}
J.~A. Laranjeira, N.~F. Martins, K.~A. Lima, L.~A. Cabral, L.~A. Ribeiro, D.~S. Galv{\~a}o, J.~R. Sambrano, Tphe-graphene: A first-principles study of a new 2d carbon allotrope for hydrogen storage, arXiv preprint arXiv:2506.00609 (2025).

\bibitem{laranjeira2025potassium}
J.~A. Laranjeira, N.~F. Martins, K.~A. Lima, B.~Aparicio-Huacarpuma, L.~A.~R. Junior, X.~Chen, D.~S. Galvao, J.~R. Sambrano, et~al., Potassium decoration on graphenyldiene monolayer for advanced reversible hydrogen storage, arXiv preprint arXiv:2506.00604 (2025).

\bibitem{laranjeira2025oli3}
J.~A. Laranjeira, W.~Elaggoune, N.~F. Martins, X.~Chen, J.~R. Sambrano, Oli3-decorated irida-graphene for high-capacity hydrogen storage: A first-principles study, arXiv preprint arXiv:2506.02375 (2025).

\bibitem{vaidyanathan2025strain}
A.~Vaidyanathan, V.~Wagh, B.~Chakraborty, A strain-engineering approach to enhance hydrogen storage in 2d holey graphyne, International Journal of Hydrogen Energy 125 (2025) 266--276.

\bibitem{wei2025rc14}
Y.~Wei, B.~Yang, S.~Zhang, H.~Chen, rc14: engineering a new dual-function carbon allotrope for sustainable energy technologies under conventional and micro-strain conditions, Journal of Materials Chemistry A (2025).

\bibitem{mohajeri2018light}
A.~Mohajeri, A.~Shahsavar, Light metal decoration on nitrogen/sulfur codoped graphyne: an efficient strategy for designing hydrogen storage media, Physica E: Low-dimensional Systems and Nanostructures 101 (2018) 167--173.

\bibitem{rowsell2005strategies}
J.~L. Rowsell, O.~M. Yaghi, Strategies for hydrogen storage in metal--organic frameworks, Angewandte Chemie International Edition 44~(30) (2005) 4670--4679.

\bibitem{shi2021high}
X.~Shi, S.~Li, J.~Li, T.~Ouyang, C.~Zhang, C.~Tang, C.~He, J.~Zhong, High-throughput screening of two-dimensional planar sp2 carbon space associated with a labeled quotient graph, The Journal of Physical Chemistry Letters 12~(47) (2021) 11511--11519.

\bibitem{gong2020theoretical}
Z.~Gong, X.~Shi, J.~Li, S.~Li, C.~He, T.~Ouyang, C.~Zhang, C.~Tang, J.~Zhong, Theoretical prediction of low-energy stone-wales graphene with an intrinsic type-iii dirac cone, Physical Review B 101~(15) (2020) 155427.

\bibitem{yin2019stone}
H.~Yin, X.~Shi, C.~He, M.~Martinez-Canales, J.~Li, C.~J. Pickard, C.~Tang, T.~Ouyang, C.~Zhang, J.~Zhong, Stone-wales graphene: A two-dimensional carbon semimetal with magic stability, Physical Review B 99~(4) (2019) 041405.

\bibitem{he2025isolated}
C.~He, S.~Li, Y.~Zhang, Z.~Fu, J.~Li, J.~Zhong, Isolated zero-energy flat bands and intrinsic magnetism in carbon monolayers, Physical Review B 111~(8) (2025) L081404.

\bibitem{Kresse_13115_1993}
G.~Kresse, J.~Hafner, \href{https://doi.org/10.1103/physrevb.48.13115}{\textit{Ab initio} molecular dynamics for open-shell transition metals}, Phys. Rev. B 48~(17) (1993) 13115--13118.
\newblock \href {https://doi.org/10.1103/physrevb.48.13115} {\path{doi:10.1103/physrevb.48.13115}}.
\newline\urlprefix\url{https://doi.org/10.1103/physrevb.48.13115}

\bibitem{Kresse_11169_1996}
G.~Kresse, J.~Furthm{\"u}ller, \href{https://doi.org/10.1103/physrevb.54.11169}{Efficient iterative schemes for \textit{Ab Initio} total-energy calculations using a plane-wave basis set}, Phys. Rev. B 54~(16) (1996) 11169--11186.
\newblock \href {https://doi.org/10.1103/physrevb.54.11169} {\path{doi:10.1103/physrevb.54.11169}}.
\newline\urlprefix\url{https://doi.org/10.1103/physrevb.54.11169}

\bibitem{Perdew1996}
J.~P. Perdew, K.~Burke, M.~Ernzerhof, \href{http://dx.doi.org/10.1103/PhysRevLett.77.3865}{Generalized gradient approximation made simple}, Physical Review Letters 77~(18) (1996) 3865–3868.
\newblock \href {https://doi.org/10.1103/physrevlett.77.3865} {\path{doi:10.1103/physrevlett.77.3865}}.
\newline\urlprefix\url{http://dx.doi.org/10.1103/PhysRevLett.77.3865}

\bibitem{perdew_1_1991}
J.~P. Perdew, \href{http://dx.doi.org/10.1016/0921-4526(91)90409-8}{Generalized gradient approximations for exchange and correlation: A look backward and forward}, Physica B: Condensed Matter 172~(1-2) (1991) 1--6.
\newblock \href {https://doi.org/10.1016/0921-4526(91)90409-8} {\path{doi:10.1016/0921-4526(91)90409-8}}.
\newline\urlprefix\url{http://dx.doi.org/10.1016/0921-4526(91)90409-8}

\bibitem{Blchl1994}
P.~E. Bl\"{o}chl, \href{http://dx.doi.org/10.1103/PhysRevB.50.17953}{Projector augmented-wave method}, Physical Review B 50~(24) (1994) 17953–17979.
\newblock \href {https://doi.org/10.1103/physrevb.50.17953} {\path{doi:10.1103/physrevb.50.17953}}.
\newline\urlprefix\url{http://dx.doi.org/10.1103/PhysRevB.50.17953}

\bibitem{Heyd2006}
J.~Heyd, G.~E. Scuseria, M.~Ernzerhof, \href{http://dx.doi.org/10.1063/1.2204597}{Erratum: “hybrid functionals based on a screened coulomb potential” [j. chem. phys. 118, 8207 (2003)]}, The Journal of Chemical Physics 124~(21) (2006) 219906.
\newblock \href {https://doi.org/10.1063/1.2204597} {\path{doi:10.1063/1.2204597}}.
\newline\urlprefix\url{http://dx.doi.org/10.1063/1.2204597}

\bibitem{Togo2023}
A.~Togo, L.~Chaput, T.~Tadano, I.~Tanaka, \href{http://dx.doi.org/10.1088/1361-648X/acd831}{Implementation strategies in phonopy and phono3py}, Journal of Physics: Condensed Matter 35~(35) (2023) 353001.
\newblock \href {https://doi.org/10.1088/1361-648x/acd831} {\path{doi:10.1088/1361-648x/acd831}}.
\newline\urlprefix\url{http://dx.doi.org/10.1088/1361-648X/acd831}

\bibitem{Grimme2006}
S.~Grimme, \href{http://dx.doi.org/10.1002/jcc.20495}{Semiempirical gga‐type density functional constructed with a long‐range dispersion correction}, Journal of Computational Chemistry 27~(15) (2006) 1787–1799.
\newblock \href {https://doi.org/10.1002/jcc.20495} {\path{doi:10.1002/jcc.20495}}.
\newline\urlprefix\url{http://dx.doi.org/10.1002/jcc.20495}

\bibitem{Hoover1985}
W.~G. Hoover, \href{http://dx.doi.org/10.1103/PhysRevA.31.1695}{Canonical dynamics: Equilibrium phase-space distributions}, Physical Review A 31~(3) (1985) 1695–1697.
\newblock \href {https://doi.org/10.1103/physreva.31.1695} {\path{doi:10.1103/physreva.31.1695}}.
\newline\urlprefix\url{http://dx.doi.org/10.1103/PhysRevA.31.1695}

\bibitem{Toh2020}
C.-T. Toh, H.~Zhang, J.~Lin, A.~S. Mayorov, Y.-P. Wang, C.~M. Orofeo, D.~B. Ferry, H.~Andersen, N.~Kakenov, Z.~Guo, I.~H. Abidi, H.~Sims, K.~Suenaga, S.~T. Pantelides, B.~\"{O}zyilmaz, \href{http://dx.doi.org/10.1038/s41586-019-1871-2}{Synthesis and properties of free-standing monolayer amorphous carbon}, Nature 577~(7789) (2020) 199–203.
\newblock \href {https://doi.org/10.1038/s41586-019-1871-2} {\path{doi:10.1038/s41586-019-1871-2}}.
\newline\urlprefix\url{http://dx.doi.org/10.1038/s41586-019-1871-2}

\bibitem{Fan2021}
Q.~Fan, L.~Yan, M.~W. Tripp, O.~Krejčí, S.~Dimosthenous, S.~R. Kachel, M.~Chen, A.~S. Foster, U.~Koert, P.~Liljeroth, J.~M. Gottfried, \href{http://dx.doi.org/10.1126/science.abg4509}{Biphenylene network: A nonbenzenoid carbon allotrope}, Science 372~(6544) (2021) 852–856.
\newblock \href {https://doi.org/10.1126/science.abg4509} {\path{doi:10.1126/science.abg4509}}.
\newline\urlprefix\url{http://dx.doi.org/10.1126/science.abg4509}

\bibitem{Liu2022}
X.~Liu, S.~M. Cho, S.~Lin, Z.~Chen, W.~Choi, Y.-M. Kim, E.~Yun, E.~H. Baek, D.~H. Ryu, H.~Lee, \href{http://dx.doi.org/10.1016/j.matt.2022.04.033}{Constructing two-dimensional holey graphyne with unusual annulative $\pi$-extension}, Matter 5~(7) (2022) 2306–2318.
\newblock \href {https://doi.org/10.1016/j.matt.2022.04.033} {\path{doi:10.1016/j.matt.2022.04.033}}.
\newline\urlprefix\url{http://dx.doi.org/10.1016/j.matt.2022.04.033}

\bibitem{Hou2022}
L.~Hou, X.~Cui, B.~Guan, S.~Wang, R.~Li, Y.~Liu, D.~Zhu, J.~Zheng, \href{http://dx.doi.org/10.1038/s41586-022-04771-5}{Synthesis of a monolayer fullerene network}, Nature 606~(7914) (2022) 507–510.
\newblock \href {https://doi.org/10.1038/s41586-022-04771-5} {\path{doi:10.1038/s41586-022-04771-5}}.
\newline\urlprefix\url{http://dx.doi.org/10.1038/s41586-022-04771-5}

\bibitem{Peng2022}
B.~Peng, \href{http://dx.doi.org/10.1021/jacs.2c08054}{Monolayer fullerene networks as photocatalysts for overall water splitting}, Journal of the American Chemical Society 144~(43) (2022) 19921–19931.
\newblock \href {https://doi.org/10.1021/jacs.2c08054} {\path{doi:10.1021/jacs.2c08054}}.
\newline\urlprefix\url{http://dx.doi.org/10.1021/jacs.2c08054}

\bibitem{Jia2017}
Z.~Jia, Y.~Li, Z.~Zuo, H.~Liu, C.~Huang, Y.~Li, \href{http://dx.doi.org/10.1021/acs.accounts.7b00205}{Synthesis and properties of 2d carbon—graphdiyne}, Accounts of Chemical Research 50~(10) (2017) 2470–2478.
\newblock \href {https://doi.org/10.1021/acs.accounts.7b00205} {\path{doi:10.1021/acs.accounts.7b00205}}.
\newline\urlprefix\url{http://dx.doi.org/10.1021/acs.accounts.7b00205}

\bibitem{Lima2025}
K.~A. Lima, J.~A. Laranjeira, N.~F. Martins, J.~R. Sambrano, A.~C. Dias, D.~S. Galvão, L.~A.~R. Junior, \href{http://dx.doi.org/10.1016/j.est.2025.117868}{Athos-graphene: Computational discovery of an art-inspired 2d carbon anode for lithium-ion batteries}, Journal of Energy Storage 133 (2025) 117868.
\newblock \href {https://doi.org/10.1016/j.est.2025.117868} {\path{doi:10.1016/j.est.2025.117868}}.
\newline\urlprefix\url{http://dx.doi.org/10.1016/j.est.2025.117868}

\bibitem{PereiraJnior2023}
M.~Pereira~Júnior, W.~da~Cunha, W.~Giozza, R.~de~Sousa~Junior, L.~Ribeiro~Junior, \href{http://dx.doi.org/10.1016/j.flatc.2023.100469}{Irida-graphene: A new 2d carbon allotrope}, FlatChem 37 (2023) 100469.
\newblock \href {https://doi.org/10.1016/j.flatc.2023.100469} {\path{doi:10.1016/j.flatc.2023.100469}}.
\newline\urlprefix\url{http://dx.doi.org/10.1016/j.flatc.2023.100469}

\bibitem{Mouhat2014}
F.~Mouhat, F.-X. Coudert, \href{http://dx.doi.org/10.1103/PhysRevB.90.224104}{Necessary and sufficient elastic stability conditions in various crystal systems}, Physical Review B 90~(22) (Dec. 2014).
\newblock \href {https://doi.org/10.1103/physrevb.90.224104} {\path{doi:10.1103/physrevb.90.224104}}.
\newline\urlprefix\url{http://dx.doi.org/10.1103/PhysRevB.90.224104}

\bibitem{Polyakova2024}
P.~V. Polyakova, R.~T. Murzaev, D.~S. Lisovenko, J.~A. Baimova, \href{http://dx.doi.org/10.1016/j.commatsci.2024.113171}{Elastic constants of graphane, graphyne, and graphdiyne}, Computational Materials Science 244 (2024) 113171.
\newblock \href {https://doi.org/10.1016/j.commatsci.2024.113171} {\path{doi:10.1016/j.commatsci.2024.113171}}.
\newline\urlprefix\url{http://dx.doi.org/10.1016/j.commatsci.2024.113171}

\bibitem{Luo2021}
Y.~Luo, C.~Ren, Y.~Xu, J.~Yu, S.~Wang, M.~Sun, \href{http://dx.doi.org/10.1038/s41598-021-98261-9}{A first principles investigation on the structural, mechanical, electronic, and catalytic properties of biphenylene}, Scientific Reports 11~(1) (Sep. 2021).
\newblock \href {https://doi.org/10.1038/s41598-021-98261-9} {\path{doi:10.1038/s41598-021-98261-9}}.
\newline\urlprefix\url{http://dx.doi.org/10.1038/s41598-021-98261-9}

\bibitem{Cheng2025}
J.~Cheng, S.~Zhou, W.~Liu, \href{http://dx.doi.org/10.1016/j.commatsci.2025.113841}{I212121 carbon: An orthorhombic carbon allotrope with superhard properties}, Computational Materials Science 253 (2025) 113841.
\newblock \href {https://doi.org/10.1016/j.commatsci.2025.113841} {\path{doi:10.1016/j.commatsci.2025.113841}}.
\newline\urlprefix\url{http://dx.doi.org/10.1016/j.commatsci.2025.113841}

\bibitem{Martins2025}
N.~F. Martins, J.~A. Laranjeira, K.~A. Lima, L.~A. Cabral, L.~A. Ribeiro, J.~R. Sambrano, \href{http://dx.doi.org/10.1016/j.apsusc.2025.163737}{Hop-graphene: A high-capacity anode for li/na-ion batteries unveiled by first-principles calculations}, Applied Surface Science 710 (2025) 163737.
\newblock \href {https://doi.org/10.1016/j.apsusc.2025.163737} {\path{doi:10.1016/j.apsusc.2025.163737}}.
\newline\urlprefix\url{http://dx.doi.org/10.1016/j.apsusc.2025.163737}

\bibitem{Alhameedi2019}
K.~Alhameedi, A.~Karton, D.~Jayatilaka, T.~Hussain, \href{http://dx.doi.org/10.1016/j.apsusc.2018.12.036}{Metal functionalized inorganic nano-sheets as promising materials for clean energy storage}, Applied Surface Science 471 (2019) 887–892.
\newblock \href {https://doi.org/10.1016/j.apsusc.2018.12.036} {\path{doi:10.1016/j.apsusc.2018.12.036}}.
\newline\urlprefix\url{http://dx.doi.org/10.1016/j.apsusc.2018.12.036}

\bibitem{Kanmani2014}
M.~Kanmani, R.~Lavanya, D.~Silambarasan, K.~Iyakutti, V.~Vasu, Y.~Kawazoe, \href{http://dx.doi.org/10.1016/j.ssc.2013.12.017}{First principles studies on hydrogen storage in single-walled carbon nanotube functionalized with tio2}, Solid State Communications 183 (2014) 1–7.
\newblock \href {https://doi.org/10.1016/j.ssc.2013.12.017} {\path{doi:10.1016/j.ssc.2013.12.017}}.
\newline\urlprefix\url{http://dx.doi.org/10.1016/j.ssc.2013.12.017}

\end{thebibliography}

\end{document}